\documentclass[aps,pra,showpacs,twoside,twocolumn,10pt]{revtex4-1}

\usepackage{amsmath,amstext,amssymb,graphicx,bm,dcolumn,color,times, braket}
\usepackage[colorlinks=true,citecolor=blue,urlcolor=blue]{hyperref}
\usepackage[table]{xcolor}

\usepackage{graphicx}

\usepackage[normalem]{ulem}

\begin{document}

\title{Multipartite entanglement at dynamical quantum phase transitions \\  with non-uniformly spaced criticalities}

\author{Stav Haldar\(^{1,2}\), Saptarshi Roy\(^1\), Titas Chanda\(^{3}\), Aditi Sen(De)\(^1\), Ujjwal Sen\(^1\)}
\affiliation{\(^1\) Harish-Chandra Research Institute, HBNI, Chhatnag Road, Jhunsi, Allahabad 211 019, India}
\affiliation{\(^2\)Hearne Institute for Theoretical Physics and Department of Physics and Astronomy,
Louisiana State University, Baton Rouge, LA USA}
\affiliation{\(^3\) Instytut Fizyki Teoretycznej, Uniwersytet Jagiello\'nski,  \L{}ojasiewicza 11, 30-348 Krak\'ow, Poland.}


\begin{abstract}

We report dynamical quantum phase transition portrait in the alternating
field transverse XY spin chain with Dzyaloshinskii-Moriya interaction by
investigating singularities in the Loschmidt echo and the corresponding
rate function after a sudden quench of system parameters. Unlike the Ising
model, the analysis of Loschmidt echo yields non-uniformly
spaced transition times in this model.  Comparative study between the
equilibrium and the dynamical quantum phase transitions in this case
reveals that there are quenches where one occurs without the other, and
the regimes where they co-exist. However, such transitions happen only when quenching is performed across at least a single gapless  or critical line. Contrary to equilibrium phase transitions,
bipartite entanglement measures do not turn out to be useful for the
detection,  while multipartite entanglement emerges as a good identifier of
this transition when the quench is done from a disordered
phase of this model.


\end{abstract}

\maketitle

\section{Introduction}
\label{sec:intro} 
Quantum many-body systems can undergo phase transitions due to a variation in the system parameters at temperatures very close to absolute zero -- entirely driven by quantum fluctuations \cite{qptbook1}. Typically, quantum critical points (QCPs)  are identified by a vanishing energy gap and divergence in characteristic correlation lengths, thereby leading to singularities in physical quantities \cite{qptbook2,qptbook3}. In recent years, bipartite as well as multipartite entanglement \cite{Horodecki2009} have been proposed to be detectors of quantum phase transitions \cite{Amico2008, Lewenstein2007, Osborne2002}.
Traditional or classical phase transitions (CPTs) are qualitatively different from quantum ones since CPTs are induced by thermal fluctuations. 
   
Apart from transitions in equilibrium, quantum systems can display non-analyticities during dynamics, a phenomenon coined as dynamical quantum phase transition (DQPT) \cite{heyl_prl, heyl_review, ASD2005, KS2004, KS2008, KS2016}, and is traditionally detected by singularities of a certain distance function  between the initial and the final time-evolved states, known as Loschmidt echo \cite{Peres}.
  Extensive DQPT studies have been performed in one-dimensional quantum spin models like XY \cite{heyl_prl, heyl_review, heyl_prl_2, heyl_prl_3, heyl_cont_sym_break, heylprl4, Vajna_prb, heylprb2, heylprb3} and XXZ \cite{XXZ} models under different  types of quenches \cite{control_dqpt}, and several counter-intuitive results have been reported regarding the relation between the equilibrium quantum phase transition (EQPT) and DQPT \cite{Vajna_prb, XXZ}. 
More importantly, DQPTs have been experimentally observed in trapped ions \cite{exp1} 
and in fermionic systems in a hexagonal lattice undergoing a topological DQPT \cite{exp2} (see also \cite{exp3}). 
  It is as yet not clear whether entanglement can be useful for detecting DQPTs. Some initial results in this direction indicate that vanishing Schmidt gap can be related to the zeroes of the Loschmidt echo \cite{schmidtgap} at critical times of the DQPT in the transverse Ising spin chain. Apart from the fundamental importance of such studies, with entanglement, it may have important implications in the design of quantum technologies like one-way and topological quantum computers and quantum simulators \cite{Bose2003, Osborne2004, Matthias2004, Raussendorf2001, Freedman2002, Georgescu2014}.


In this paper, we examine DQPT for a uniform and alternating transverse field XY spin chain (ATXY) in presence of an additional antisymmetric interaction,  the Dzyaloshinskii-Moriya (DM) interaction \cite{amadr_dm,dm_moriya1,dm_moriya2,dm_anderson,dm1,dm2,dm3,dm4,dm5,dm6,dm7,dm8,
 dm9,dm10,dm11,dm12,dm13,dm14,dm15,dm16,dm17,dm18,dm19,dm20,dm21,dm22,dm23,dm24,dm25,dm26,
 dm27,dm28,dm29,dm30} via both the traditional and information theoretic approach.
 For the DATXY model, we analytically compute the Loschmidt echo and its corresponding critical times which turn out to be non-uniformly spaced for most general quenches of the system parameters, viz. the uniform field, the alternating field, and the DM interaction strength. The quenches are performed both within and across the equilibrium phase boundaries. 
The DATXY model reduces to various well-known models like Ising, UXY, XX etc. in different limits of the system parameters, and reproduces the critical times of DQPT for the same \cite{heyl_prl}.
Such a general model is chosen because the reduced models, in the case of DQPTs, have shown not to capture the relevant physics in its full generality. For example, the Ising model predicts an equivalence between EQPT and DQPT \cite{heyl_prl}, although later this was shown not to be the case even for the UXY model \cite{Vajna_prb} (see also \cite{XXZ}).
The results for the DATXY model also confirms this inequivalence.
 Moreover, our results provide a necessary condition on the quenches that lead to a DQPT, namely: \emph{a quench corresponding to a DQPT must cross at least one equilibrium critical line.}

On the other hand, systematic studies reveal that unlike in EQPT, bipartite entanglement fails as a detector for DQPT. However, we find that if the initial state belongs to a disordered phase,  genuine multipartite entanglement \emph{can} identify a DQPT.
Specifically, we observe that the time-averaged standard deviation of a geometric measure of multipartite entanglement \cite{ggm} (see also \cite{ggm2,ggm3,ggm4,ggm5}) has \emph{much} higher values when the final quench point falls in a region that corresponds to a DQPT than in regions without it. The regions show a large overlap with those detected through singularities in the Loschmidt echo. Our observation also indicates that low value of multipartite entanglement in the initial state, which indeed is a feature of ground states in the disordered phase,  can be a plausible explanation for this asymmetric identification of DQPT with multiparty entanglement.

The paper is organized as follows. 
The model Hamiltonian, its diagonalization, and its ground state phases are discussed in Sec. \ref{sec:system}. In Sec. \ref{sec:dqpt}, the DQPT exhibited by the model is analyzed via Loschmidt echo,
 while the same is reanalyzed by the tools of quantum information theory, namely by employing bipartite and genuine multipartite entanglement in Sec. \ref{sec:ent}. We finally conclude in Sec. \ref{sec:con}.   

\section{Transverse XY model with alternating field and anti-symmetric interaction}
\label{sec:system}

We consider a paradigmatic family of interacting quantum spin-$1/2$ systems on a one-dimensional (1D) lattice with nearest-neighbor anisotropic XY interaction as well as asymmetric Dzyaloshinskii-Moriya (DM) interaction in presence of uniform and alternating external transverse magnetic fields described by the Hamiltonian \cite{amadr_dm},
\begin{eqnarray}
\hat{H} &=& 
\frac{1}{2}\sum_{j=1}^N \Big[J \big(\frac{1+\gamma}{2}\hat{\sigma}^x_j\hat{\sigma}^x_{j+1} + \frac{1-\gamma}{2}\hat{\sigma}^y_j\hat{\sigma}^y_{j+1}\big) \nonumber \\
&+& \frac{D}{2}\big(\hat{\sigma}^x_j\hat{\sigma}^y_{j+1}-\hat{\sigma}^y_j\hat{\sigma}^x_{j+1} \big)+ \big(h_1 + (-1)^j h_2\big) \hat{\sigma}^z_j\Big],
\label{eq:hamil}
\end{eqnarray}
with periodic boundary condition, i.e., $\hat{\sigma}_{N+1} = \hat{\sigma}_1$.
Here, $\hat{\sigma}^\alpha , \alpha = x, y, z$ are the Pauli matrices, $J$ and $D$ represent the strengths of nearest-neighbor exchange interaction and the DM interaction respectively,  $\gamma$ $(\neq 0)$ is the anisotropy parameter for the $xx$ and $yy$ interactions, $h_1$ and $h_2$ are the uniform and alternating transverse magnetic fields respectively, and $N$ denotes the total number of lattice sites. 
The Hamiltonian, referred to as the DATXY model, can be mapped to a spinless 1D Fermi system with two sublattices (for even and odd sites) via the Jordan-Wigner transformation \cite{amadr_dm, atxy_pra}. Further, performing Fourier transformation, the Hamiltonian can be block-diagonalized in the momentum space, as $\hat{H} = \sum_{p=1}^{N/4}\hat{H}_p$, with 
\begin{eqnarray}
\hat{H}_p &=& J\Big[(\cos\phi_p + d\sin\phi_p)(a_p^\dagger b_p + b_p^\dagger a_p) \nonumber \\ 
&+&(\cos\phi_p - d\sin\phi_p)(a_{-p}^\dagger b_{-p} + b_{-p}^\dagger a_{-p}) \nonumber \\ 
&-& i \gamma \sin\phi_p(a_p^\dagger b_{-p} + a_p b_{-p} - a_{-p}^\dagger b_{p}^\dagger - a_{-p} b_p) \nonumber \\
&+& (\lambda_1 + \lambda_2) (b_p^\dagger b_p + b_{-p}^\dagger b_{-p})  \nonumber \\
&+&  (\lambda_1-\lambda_2) (b_p^\dagger b_p + b_{-p}^\dagger b_{-p})-2\lambda_1\Big], 
\label{eq:fermi_hamil}
\end{eqnarray}
where $\lambda_i= h_i/J$ with $i=1,2$ and $d = D/J$  are the dimensionless system parameters, $\phi_p = 2 \pi p /N$, and $\hat{a}_p$ ($\hat{b}_p$) corresponds to the fermionic operators for odd (even) sublattices. Therefore, diagonalization of $\hat{H}$, required to study its characteristics, reduces in diagonalizing $\hat{H}_p$ for different momentum sectors, which can be done by proper choice of the basis \cite{amadr_dm, atxy_pra}. In equilibrium, this model can exhibit two paramagnetic phases (PM-I and PM-II), an antiferromagnetic phase (AFM) and a \emph{gapless} chiral (CH) phase (see \cite{amadr_dm}).

It is noteworthy to mention that  several well-known quantum spin models in different parameter regimes can be obtained from the DATXY model, such as
\begin{enumerate}
\item transverse field Ising (TFI) model for $\gamma=1$, $\lambda_2 = d = 0$,
\item quantum XY model with uniform magnetic field (UXY)  for $\lambda_2 = d = 0$,
\item quantum XY model with uniform and alternating magnetic fields (ATXY) \cite{qptbook3, atxy_1, atxy_2, atxy_3, atxy_pra}  for $d = 0$, and
\item quantum XY model with uniform magnetic field in presence of DM interaction (DUXY) for $\lambda_2 = 0$.
\end{enumerate}

\begin{figure}
\includegraphics[width=\columnwidth]{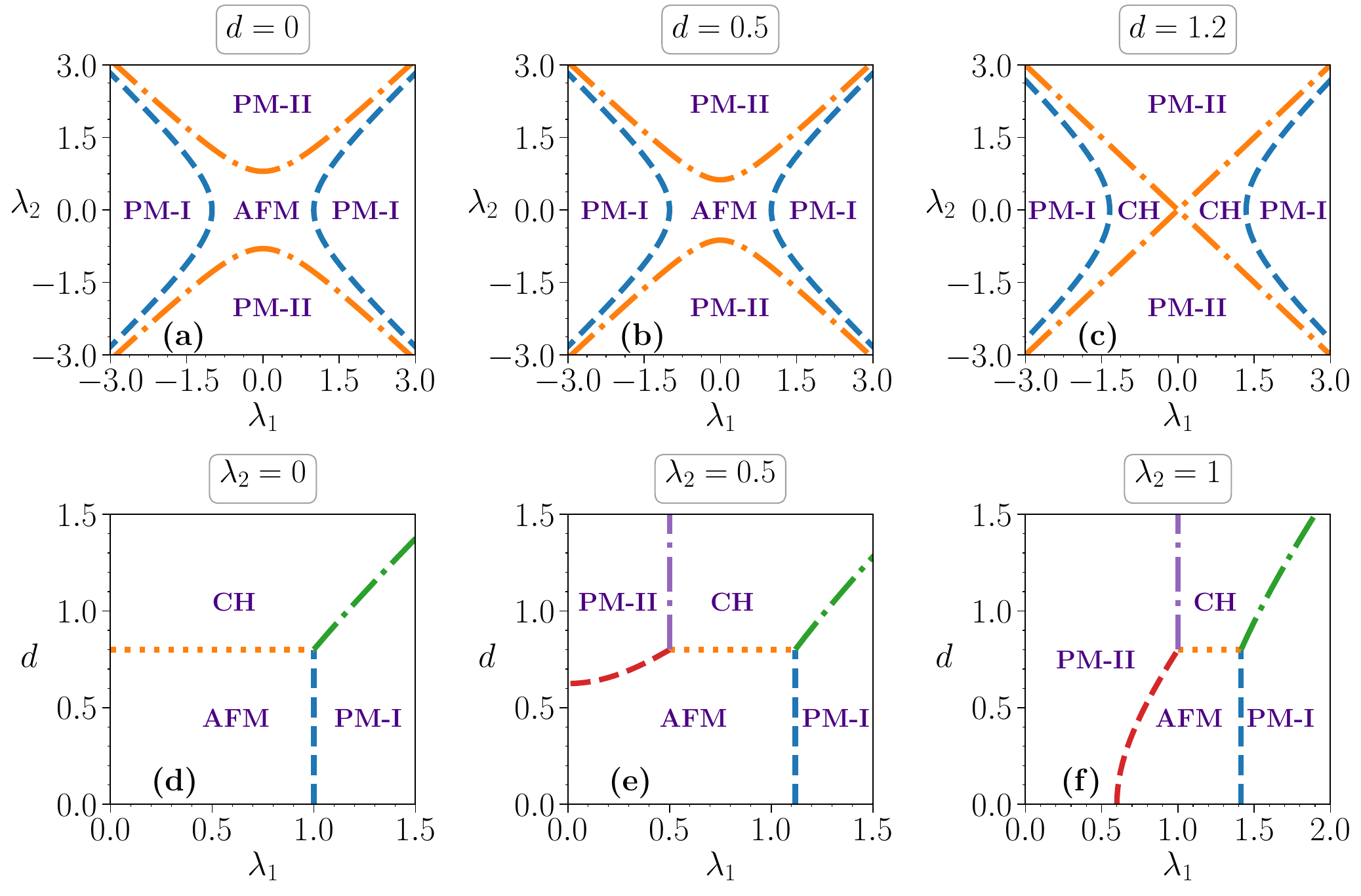}
\caption{(Color online.) Phase diagrams of the DATXY model in different parameter-spaces. Here, we choose $\gamma = 0.8$. Unless otherwise stated, we will use $\gamma = 0.8$ throughout the paper for demonstration purpose. All quantities plotted are dimensionless.}
\label{fig:phase}
\end{figure}

We choose the DATXY model for demonstration, as this model possesses a very rich phase diagram at zero temperature (see Fig. \ref{fig:phase}).
Moreover, we notice that by fixing different parameters suitably,  it can be reduced to any of the above four models.  
The equilibrium quantum phase transitions between these different phases occur across the following surfaces \cite{amadr_dm}:
\begin{itemize}
\item For\hspace{0.3 cm}$0 \leq d<\gamma$,
  \begin{enumerate}
  \item $\lambda^2_1 = 1 + \lambda^2_2$ \hspace{1.25 cm}(PM-I $\leftrightarrow$ AFM)
  \item $\lambda^2_2 = \lambda^2_1 + \gamma^2 - d^2$ \hspace{0.3 cm}(PM-II $\leftrightarrow$ AFM)
  \end{enumerate}
  \item For\hspace{0.3 cm}$d>\gamma$,
  \begin{enumerate}
  \item $\lambda^2_1 = 1 + \lambda^2_2 + d^2 - \gamma^2$  \hspace{0.25 cm}(PM-I $\leftrightarrow$ CH)
  \item $\lambda_1 = \pm \lambda_2$ \hspace{2.15 cm}(PM-II $\leftrightarrow$ CH).
  \end{enumerate}
\end{itemize} 
Note that, in the thermodynamic limit, the AFM phase of the spin Hamiltonian given in Eq. \eqref{eq:hamil} has two-fold degeneracy whereas, in the fermionic version of the model, the ground state in that phase is unique. The AFM phase appears for $d<\gamma$, whereas for $d>\gamma$, we get the CH phase. The gapless CH phase, in addition of having a continuous spectra, has three-fold degenerate ground state.   

\section{Dynamical quantum phase transitions in DATXY model}
\label{sec:dqpt} 
Let us now move to investigate dynamical quantum phase transitions (DQPT) in the DATXY model. 
At $t=0$, we prepare the system in a ground state of a Hamiltonian, $\hat{H}^{(0)} = \hat{H}(g_0)$, with initial parameter values, $g_0 \equiv \{\lambda_1(t=0), \lambda_2(t=0), d(t=0)\}$, and then at  $t>0$, we suddenly quench the system parameters to new values, $g_1 \equiv \{\lambda_1(t>0), \lambda_2(t>0), d(t>0)\}$, such that the new Hamiltonian becomes $\hat{H}^{(1)} = \hat{H}(g_1)$, according to which the system evolves  with time. Note that, unless otherwise stated, we will not change the anisotropy parameter, $\gamma$, in the quenching process.

\subsection{Loschmidt amplitude, Loschmidt echo}
\label{sec:LE}
Analogous to the role of canonical partition function in temperature driven phase transitions, the Loschmidt amplitude is shown to play an important role in  DQPTs \cite{heyl_prl, heyl_review}, and is defined as the overlap of the time evolved state of a system with its initial state. If the initial state, $\ket{\Psi^0}$, is prepared as the ground state of the initial Hamiltonian, $\hat{H}^{(0)}$, and the Hamiltonian after the  quench is $\hat{H}^{(1)}$, then the  Loschmidt amplitude is defined as
\begin{eqnarray}
G(t)=\bra{\Psi^0}e^{-i \hat{H}^{(1)} t / \hbar} \ket{\Psi^0}.
\label{eq:LA}
\end{eqnarray}  
For quenching in the parameter-space of the DATXY model, using Eq. \eqref{eq:fermi_hamil}, the above expression can be decomposed as
\begin{eqnarray}
G(t) = \prod_{p = 1}^{N/4} \bra{\Psi^0_p}e^{-i \hat{H}^{(1)}_p t / \hbar} \ket{\Psi^0_p} = \prod_{p=1}^{N/4}G_p(t),
\end{eqnarray} 
where $\ket{\Psi^0_p}$ is the eigenstate of $\hat{H}_p^{(0)}$ corresponding to the lowest eigenvalue, and in the last expression, we have defined Loschmidt amplitude per momentum mode as $G_p(t) = \bra{\Psi^0_p}e^{-i \hat{H}^{(1)}_p t / \hbar} \ket{\Psi^0_p}$.
The Loschmidt echo, $L(t)$, is then described by the probability associated with this amplitude, i.e., $L(t)= |G(t)|^2$. The rate function associated with $L(t)$, which is analogous to the free energy (per lattice-site) in thermal phase transitions, can be defined as 
\begin{eqnarray}
\mathcal{F}(t) &=& -\lim_{N\rightarrow \infty} \frac{1}{N} \log L(t) = -\lim_{N \rightarrow \infty} \frac{2}{N} \sum_{p=1}^{N/4} \log |G_p(t)| \nonumber \\
\label{eq:rf}
\end{eqnarray} 
Similar to the thermal phase transition, where transition is dictated by the nonanalytic behavior of the associated free energy with respect to the temperature, the DQPT can be detected by the nonanalyticity of the rate function as a function of time at some critical time $t^*$.

To deduce the analytical expressions of the Loschmidt amplitude and the rate function for a general quench from $g_0 \equiv \{\lambda_1(t=0),\lambda_2(t=0),d(t=0)\}$ to  $g_1 \equiv \{\lambda_1(t>0),\lambda_2(t>0),d(t>0)\}$, we introduce the fermionic vector operator $\hat{A_p} = \begin{bmatrix} \hat{a}_p \\ \hat{b}_p  \end{bmatrix}$. We then perform a Bogoliubov transformation of the following form:
\begin{eqnarray}
\begin{bmatrix}
\hat{A}_p \\ \hat{A}^\dagger_{-p}
\end{bmatrix} = 
M_p  \begin{bmatrix} 
    \hat{\Gamma}_p\\
    \hat{\Gamma}^\dagger_{-p}
    \end{bmatrix}
= \begin{bmatrix} 
    U_p & -iV_p\\
    -iV^*_p & U^*_p
    \end{bmatrix}  
    \begin{bmatrix} 
    \hat{\Gamma}_p\\
    \hat{\Gamma}^\dagger_{-p}
    \end{bmatrix},
\label{eq:bogol}
\end{eqnarray}
with $\hat{\Gamma}_p = \begin{bmatrix} \hat{\eta}^a_p \\ \hat{\eta}^b_p  \end{bmatrix}$, such that $\hat{H}_p$ is diagonal in the Bogoluibov basis,
$\{ {{}\hat{\eta}^a_p } ^ {\dagger}, {{}\hat{\eta}^b_p}^{\dagger}, {\hat{\eta}^a_{-p}}, {\hat{\eta}^b_{-p}} \}$. $U_p$ and $V_p$ are Bogoliubov coefficients (matrices) and are functions of $g \equiv \{\lambda_1,\lambda_2, d\} $. 

In order to calculate $\bra{\Psi^0}e^{-iH^{(1)}t/\hbar}\ket{\Psi^0}$, we express the operators, $\{\hat{\Gamma}_p(g_0)\}$ that diagonalize the initial Hamiltonian, $\hat{H}^{(0)}$, in terms of $\{\hat{\Gamma}_p(g_1)\}$ which diagonalize the final Hamiltonian, $\hat{H}^{(1)}$ using Eq. \eqref{eq:bogol} as
 \small
 \begin{eqnarray}
 \begin{bmatrix} 
    \hat{\Gamma}_p(g_0)\\
    \hat{\Gamma}^{\dagger}_{-p}(g_0)
    \end{bmatrix} = \begin{bmatrix} 
    \mathcal{U}_p(g_0, g_1) & -i\mathcal{V}_p(g_0, g_1)\\
    -i\mathcal{V}^*_p(g_0, g_1) & \mathcal{U}^*_p(g_0, g_1)
    \end{bmatrix}  \begin{bmatrix} 
    \hat{\Gamma}_p(g_1)\\
    \hat{\Gamma}^\dagger_{-p}(g_1)
    \end{bmatrix}, 
\end{eqnarray}  
\normalsize
with
 \begin{eqnarray}
    \mathcal{U}_p (g_0, g_1)&=& U^\dagger_p(g_0) U_p(g_1) + V^T_p(g_0)V^*_p(g_1), \nonumber \\
    \mathcal{V}_p (g_0, g_1)&=& U^\dagger_p(g_0)V_p(g_1) - V^T_p(g_0)U^*_p(g_1).
    \end{eqnarray}
We can now write the initial state $\ket{\Psi^0}$ as a boundary state composed of zero-momentum modes of $\hat{H}^{(1)}$, given by
\begin{eqnarray}
\ket{\Psi^0} = \mathcal{N}^{-1}\exp\Big[i\sum_ {p=1}^{N/4} \hat{\Gamma}^{\dagger T}_p (\mathcal{U}^{-1}_p\mathcal{V}_p) \hat{\Gamma}^\dagger_{-p})\Big]\ket{0},
\label{eq:bs}
\end{eqnarray}
where $\ket{0}$ is the ground state of $\hat{H}^{(1)}$, $\mathcal{N}$ is the normalization constant, and $T$ denotes the transpose of the corresponding operators.  By using eigenvalues $\hbar\omega_p^k$ $(k = 1,2,3,4$) of $\hat{H}_p^{(1)} = \hat{A}_p^{\dagger}  \tilde{H}_p^{(1)} \hat{A}_p$, where $\hat{A}_p$ is the column vector,  $(\hat{a}_p, \hat{b}_p, \hat{a}_{-p}^{\dagger}, \hat{b}_{-p}^{\dagger})$ and 
\begin{widetext}
\small
\begin{equation}
\tilde{H}_p^{(1)} = J
\begin{bmatrix}
(\lambda_1 - \lambda_2) &  (\cos\phi_p + d \sin\phi_p) & 0 & -i \gamma \sin\phi_p \\
(\cos\phi_p + d \sin\phi_p)  & (\lambda_1 + \lambda_2) & -i \gamma \sin\phi_p &0 \\
0 & i\gamma\sin\phi_p & -(\lambda_1 - \lambda_2) &  -(\cos\phi_p - d \sin\phi_p) \\
i \gamma \sin\phi_p & 0 & -(\cos\phi_p - d \sin\phi_p) & -(\lambda_1 + \lambda_2)
\end{bmatrix},
\label{eq:hp_matrix_alt}
\end{equation}
\normalsize
with $\phi_p \in [-\pi/2,\pi/2]$, we obtain
%
\begin{eqnarray}
G_p(t) &&= e^{i \frac{J t}{\hbar} (\omega_p^3 + \omega_p^4)} \nonumber \\
&&
\resizebox{0.9\hsize}{!}{$\times
\frac{1 
+ e^{-i \frac{J t}{\hbar} (\omega^1_p - \omega^3_p)}|\mathcal{T}^p_{11}|^2 
+ e^{-i \frac{J t}{\hbar} (\omega^2_p - \omega^4_p)} |\mathcal{T}^p_{22}|^2
+ e^{-i \frac{J t}{\hbar}(\omega^1_p - \omega^4_p)}|\mathcal{T}^p_{12}|^2 
+ e^{-i \frac{J t}{\hbar}(\omega^2_p - \omega^3_p) } |\mathcal{T}^p_{21}|^2 
+  e^{-i \frac{J t}{\hbar}(\omega^1_k+\omega^2_k - \omega^3_k-\omega^4_k )} |\mathcal{T}^p_{11}\mathcal{T}^p_{22} - \mathcal{T}^p_{12}\mathcal{T}^p_{21}|^2}
{1 + \vert\mathcal{T}^p_{11} \vert^2 + \vert \mathcal{T}^p_{22} \vert^2 + \vert \mathcal{T}^p_{12} \vert^2 + \vert \mathcal{T}^p_{21} \vert^2 + \vert \mathcal{T}^p_{11}\mathcal{T}^p_{22} - \mathcal{T}^p_{12}\mathcal{T}^p_{21} \vert^2},$} \nonumber \\
\label{eq:LA_p}
\end{eqnarray} 
\normalsize
\end{widetext}
with $\mathcal{T}^p_{ij} = (\mathcal{U}^{-1}_p\mathcal{V}_p)_{ij}$, see Appendix \ref{ap:nonana-ratefn} for details.
Finally, we get the rate function associated with the quench from parameters $g_0$ to $g_1$, in the thermodynamic limit, as
  \begin{eqnarray}
  \mathcal{F}(t) = -\int^{\frac{\pi}{2}}_0 \frac{d\phi_p}{\pi} \log |G_p(t)|.
  \label{eq:rfdatxy}
  \end{eqnarray}
Now, the solutions of $|G_p(t)| = 0$, correspond to the critical times, denoted by $t^*$. 
In case of the UXY model (i.e., $\lambda_2 = d = 0$), we get $\tau_{12}^p = \tau_{12}^p = 0$, making $\tau^p$-matrix diagonal.
So, the expression for $G_p(t)$ simplifies to the form,
\begin{eqnarray}
\resizebox{1\hsize}{!}{$
G_p(t) = e^{-i \frac{J t}{\hbar} (\omega^1_p + \omega^2_p)} 
\frac{(1 + e^{-2 i \frac{J t}{\hbar}\omega^1_p } |\mathcal{T}_{11}^p|^2) (1 + e^{-2 i \frac{J t}{\hbar}\omega^2_p } |\mathcal{T}_{22}^p|^2)}
{(1 + |\mathcal{T}_{11}^p|^2) (1 + |\mathcal{T}_{22}^p|^2)}.$} \nonumber \\
\end{eqnarray}
The solutions of $|G_p(t)| = 0$, lead to critical times, $t^* = \frac{\hbar\pi}{J\omega^1_{p^*}}(n+\frac{1}{2}),$ $n = 1, 2, 3...$  \cite{ASD2005}, which matches with the known results of for the TFI model \cite{heyl_prl}. The equation for $\omega_p^2$ does not give any critical point. 
In case of $\lambda_2 \neq 0$, such simplification does not happen, and therefore, for quenching onto any phases of the ATXY model (i.e., for $d$ $(t>0) = 0$),
the solutions of $|G_p(t)| = 0$ are obtained numerically by a standard root-finding algorithm, which is naturally the case for the DATXY model. For details of the calculation of rate function, see Appendix \ref{ap:nonana-ratefn}.


Note that the above treatment has to be generalized if the initial Hamiltonian, $\hat{H}^{(0)}$, has degenerate ground states (see Refs. \cite{heyl_prl, heyl_review, heyl_prl_2, heyl_prl_3, heyl_cont_sym_break}). 
 To keep things simple, we will work with the fermionic version of the model, which is free from degeneracy in the AFM phase, and will not consider CH phase as the initial one since, the above analysis has to be modified in that situation. However, the final parameters of the quench in the $(\lambda_1, \lambda_2, d)$ space, can belong to any phase.

\subsubsection{Non-uniformly spaced critical times}
\label{sec:non-uniform}
\begin{figure}[ht]
\includegraphics[width=0.7\linewidth]{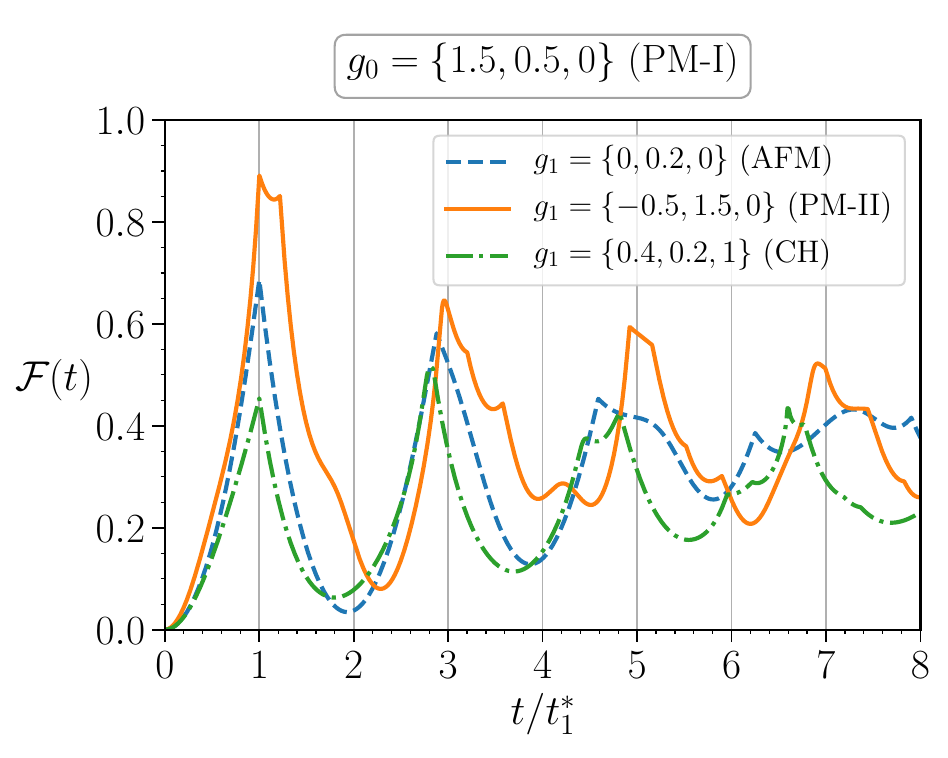}
\caption{(Color online.) Rate function, $\mathcal{F}(t)$ vs. $t/t_1^*$. The dynamics occurs due to quenches from the point $g_0 = \{1.5, 0, 0\}$, which lies in the PM-I phase, to three different points, namely $(1)$ $g_1 = \{0, 0.2, 0\}$ (AFM phase), $(2)$ $g_1 = \{-0.5, 1.5, 0\}$ (PM-II phase), and $(3)$ $g_1 = \{0.4, 0.2, 1\}$ (CH phase), in the parameter-space of the DATXY model. All the quenches are across EQPT lines. Clearly, in all the three cases, $\mathcal{F}(t)$ becomes nonanalytic for different critical times, $t^*$.  Here, we have normalized the time-axis by the first critical time $t^*_1$ in all the three cases to highlight the fact that in the DATXY model, the critical times are not uniformly spaced with each other. 
 All quantities plotted are dimensionless.}
 \label{fig:from_pm1}
\end{figure}

For the DATXY model, in general, the $\tau^p$-matrix has off-diagonal terms, which possibly leads to nonuniformly-spaced critical times $t^*$  on the time axis as depicted in Fig. \ref{fig:from_pm1}. Note that this was not the case for the quantum XY model with uniform magnetic field (UXY) or the TFI model \cite{heyl_prl}. 

\begin{figure}
\includegraphics[width=\columnwidth]{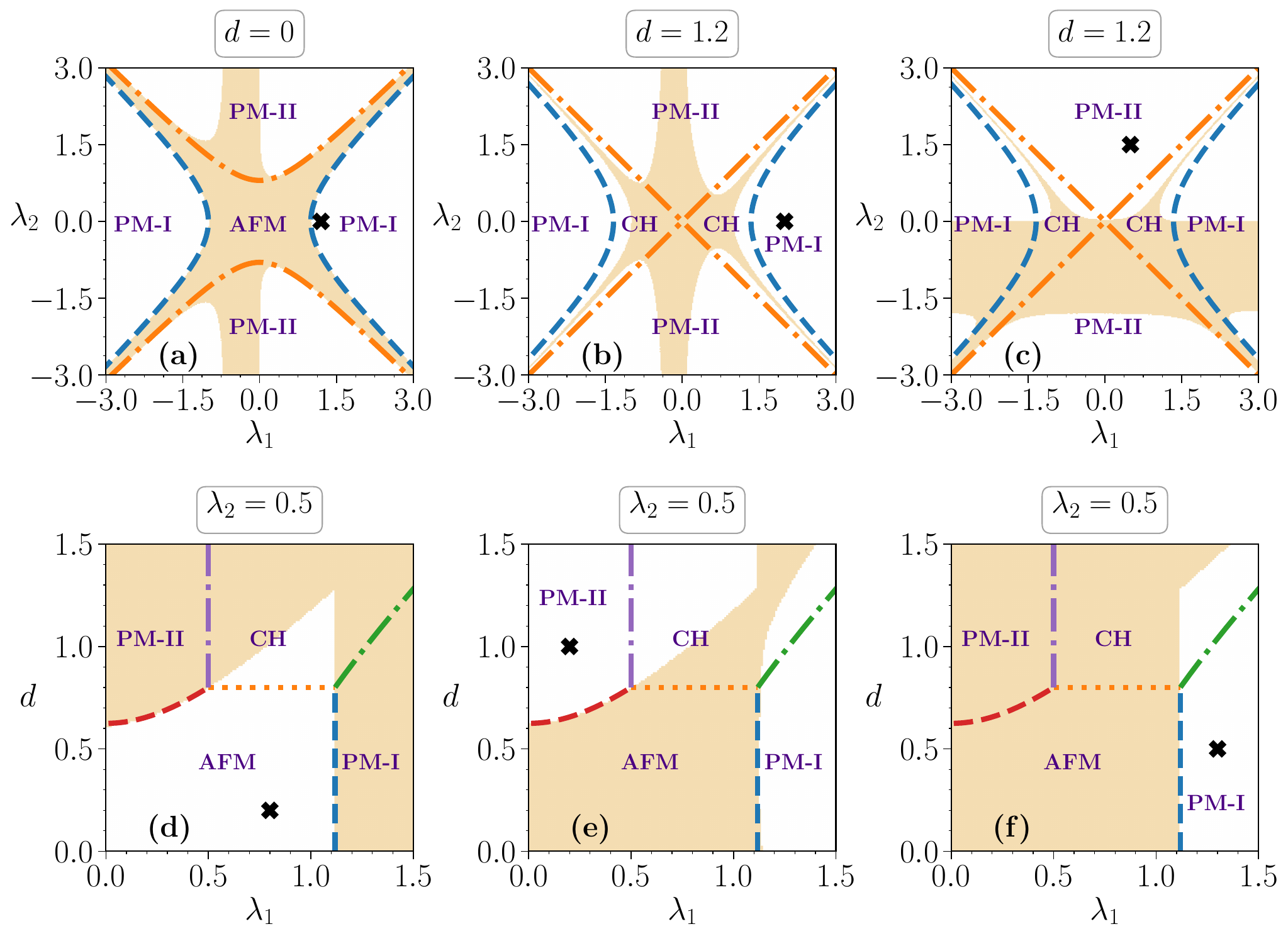}
\caption{(Color online.)  DQPT regions in the parameter-space of the DATXY model for the quenches from the points marked by ``$\bm{\times}$''. When a quantum quench is performed from a point, marked by the symbol ``$\bm{\times}$'', to any other point on the parameter space, only the shaded region shows nonanalyticity in the rate function given in Eq. \eqref{eq:rfdatxy}. 
All the axes are dimensionless.}
\label{fig:dpt_ept}
\end{figure}

\subsubsection{Connection between DQPT and EQPT}
\label{sec:connection}
For the TFI model in 1D, it was found that DQPTs   are in one-to-one correspondence to the EQPTs \cite{heyl_prl, heyl_review}, where the  nonanalytic nature of the rate function was only observed for a quench across the EQPT line. However, later on, counterexamples in the UXY model were reported in \cite{Vajna_prb}, where such connections were absent.
 In case of the DATXY model, the situation is much more involved and we summarize our observations below. 

\begin{enumerate}
\item For $d < \gamma$, and quenches from AFM phase to one of the PM phases, or vice-versa, DQPT has one-to-one correspondence with EQPT
(see Fig. \ref{fig:dpt_ept} (d)). However, if the anisotropy parameter, $\gamma$, is also quenched, such a connection no longer holds (see Ref. \cite{Vajna_prb}).

\item As the PM phases of DATXY model are connected by local transformations \cite{atxy_1}, one expects that quenching between these two phases does not result in a DQPT (see Fig. \ref{fig:dpt_ept}(e)). On the contrary, our analysis shows existence of DQPTs for some specific cases of such quenches (see Fig. \ref{fig:dpt_ept}(a)-(c), and (f)).


\item For $d > 0$, a quench from a point in PM-II to another point in PM-II with different signs in $\lambda_2$ (involving a crossing of one or more gapless critical lines/regions) may result in a DQPT (Fig. \ref{fig:dpt_ept}(c)). 
Note that this feature is special only to the PM-II phase, and can not be observed in any other phases for a fixed value of the anisotropy parameter, $\gamma$. Such observation is akin to the quench of $\gamma$ as seen in Ref. \cite{Vajna_prb}.

\item For $\lambda_2=0$ and fixed $\lambda_1$, quenching into the CH phase from the AFM phase by increasing $d$ in the quantum quench does not result into a DQPT, as, in that case, the initial Hamiltonian, $\hat{H}(\lambda_1, 0, d (< \gamma))$, and the final Hamiltonian, $\hat{H}(\lambda_1, 0, d(> \gamma))$, commute  with each other. However, with $\lambda_2 \neq 0$, a DQPT from the AFM to CH phase is possible, provided the quenching is done sufficiently deep into the CH phase (Fig. \ref{fig:dpt_ept}(d)). Similar features are observed for PM-I(II) $\rightarrow$ CH quenches (see Figs. \ref{fig:dpt_ept}(b), (c), (e) and (f)).

\end{enumerate}
From the above observations, it is clear that, typically, there is no one-to-one connection between DQPT and EQPT phases in the DATXY model. However, all the observations reported here indicate that the \emph{non-analyticity of the rate function implies quench across at least a single critical line,} thereby providing us with a \emph{necessary} condition for obtaining a DQPT.
Now, we would analyze DQPT from an information-theoretic point of view 
and compare it with the Loschmidt echo-based approach.

\section{Entanglement as a potential detector of DQPT}
\label{sec:ent}
In the case of equilibrium quantum phase transitions (EQPTs), entanglement, both bipartite as well as multipartite, emerges  as efficient detectors \cite{qpt-ent, qptbook1}. 
It was even shown that in some models, where the traditional detection methods fail, EQPT could be detected by entanglement-based quantities \cite{frankver, AKLT}.
Later, in other works \cite{Biswas2014,Chha2007,manab-ggm}, it was found that there exist models for which multipartite entanglement turns out to be better for identifying EQPT compared to bipartite measures. In general, it has been realized that  entanglement as well as other quantum correlation measures  have the potential to detect EQPTs. The question then is -- \emph{Can entanglement be a ``good" quantity to identify phase transitions that occur with the variation of time, after a sudden quench of parameters?}

We answer this question by analyzing the time evolution of bipartite and multipartite entanglement for the DATXY model after a sudden quench. Our analyses reveal, bipartite entanglement, contrary to its efficient EQPT detection capability for this model, turns out to be an inefficient detector of DQPT. 
First of all, the failure of bipartite entanglement as an identifier of DQPT for this model already once more underlines the difference between EQPT and DQPT from an information-theoretic perspective.
Moreover, multipartite entanglement emerges as a good detector of DQPT. 

\subsection{Bipartite Entanglement fails to detect DQPT}
In this paper, we quantify bipartite entanglement via logarithmic-negativity, $\mathcal{L}$.
For an arbitrary two party density matrix, $\rho_{AB}$, negativity ($\mathcal{N}$) and $\mathcal{L}$ are defined as
\begin{eqnarray}
\mathcal{N}(\rho_{AB}) &=& \frac{1}{2}(||\rho^{T_B}_{AB}|| - 1) = \frac{1}{2}(||\rho^{T_A}_{AB}|| - 1), \nonumber \\ 
\mathcal{L}(\rho_{AB}) &=& \log_2(2\mathcal{N}(\rho_{AB})+1),
\end{eqnarray}
where $||A|| = \text{tr}\sqrt{A^\dagger A}$ and $T_{A(B)}$ in the superscript of $\rho_{AB}$ denotes partial transposition in party $A(B)$. Note that for $2\otimes2$ and $2\otimes3$ systems, negative partial transposition and hence non-zero $\mathcal{L}$ provides a necessary and sufficient condition for guaranteeing entanglement \cite{necess-suff}. Thus in our case, since all the two-site reduced density matrices have dimension $2\otimes2$, $\mathcal{L}$ is a faithful measure of entanglement.

Our analysis establishes that nearest neighbor entanglement shows some qualitative changes when a quench is performed across a disorder to order transition, i.e. PM-I (II) $\rightarrow$ AFM/CH phase. 
 Specifically, in these cases, the dynamics of $\mathcal{L}$ displays a distinctive \emph{collapse and revival} feature.
On the other hand, if the final parameters of the quench correspond to a disordered phase which is same as the phase of the initial state, $\mathcal{L}$ does not show any collapse or revival and simply oscillates with decreasing amplitude, finally reaching a steady value.

However, note that the above features are only general trends and there exist several counter-examples to these patterns. Further investigation reveals that the dynamics of $\mathcal{L}$  shows a large overlap with the equilibrium phases and only has a weak connection with DQPT. Hence, we infer that bipartite entanglement is not an efficient detector of DQPT.

\subsection{Effective definition of the generalized geometric measure (GGM)}

Before presenting the results, let us define the entanglement measure that we use to study multipartite entanglement.
 For a set of states which are non-genuinely multipartite entangled, denoted by $nG$, the GGM of a state $|\psi\rangle$, is defined by
\begin{eqnarray}
\mathcal{G}(|\psi\rangle) = 1-\max \vert \langle\phi\vert\psi\rangle\vert^2, \ket{\phi} \in nG, 
\end{eqnarray}
which, for a $N$-party pure state, reduces to
\begin{eqnarray}
\mathcal{G}(|\psi\rangle) = 1 &-& \max \lbrace \mu^{\max}_{i_1:\text{rest}},\mu^{\max}_{i_1i_2:\text{rest}},..., \mu^{\max}_{i_1i_2....i_{M}:\text{rest}}|\nonumber \\
i_1,i_2...i_{M} &\in& \{1,2,... \tilde{N}\}; i_k \neq i_l; k,l \in \{1,2,... M\} \rbrace, \nonumber \\ 
\end{eqnarray}
where $\tilde{N} = N/2$ or $(N-1)/2$ for even and odd lattice sizes respectively, and $\mu^{\max}$ denotes the maximal eigenvalue of the reduced density matrices with rank equal to the number of $i$'s present in the subscript of $\mu$. Therefore, the evaluation of GGM, $\mathcal{G}(|\psi\rangle)$, boils down to the evaluation of the maximum of maximal eigenvalues for all reduced density matrices.

\begin{figure}[h]
\includegraphics[width=0.9\linewidth]{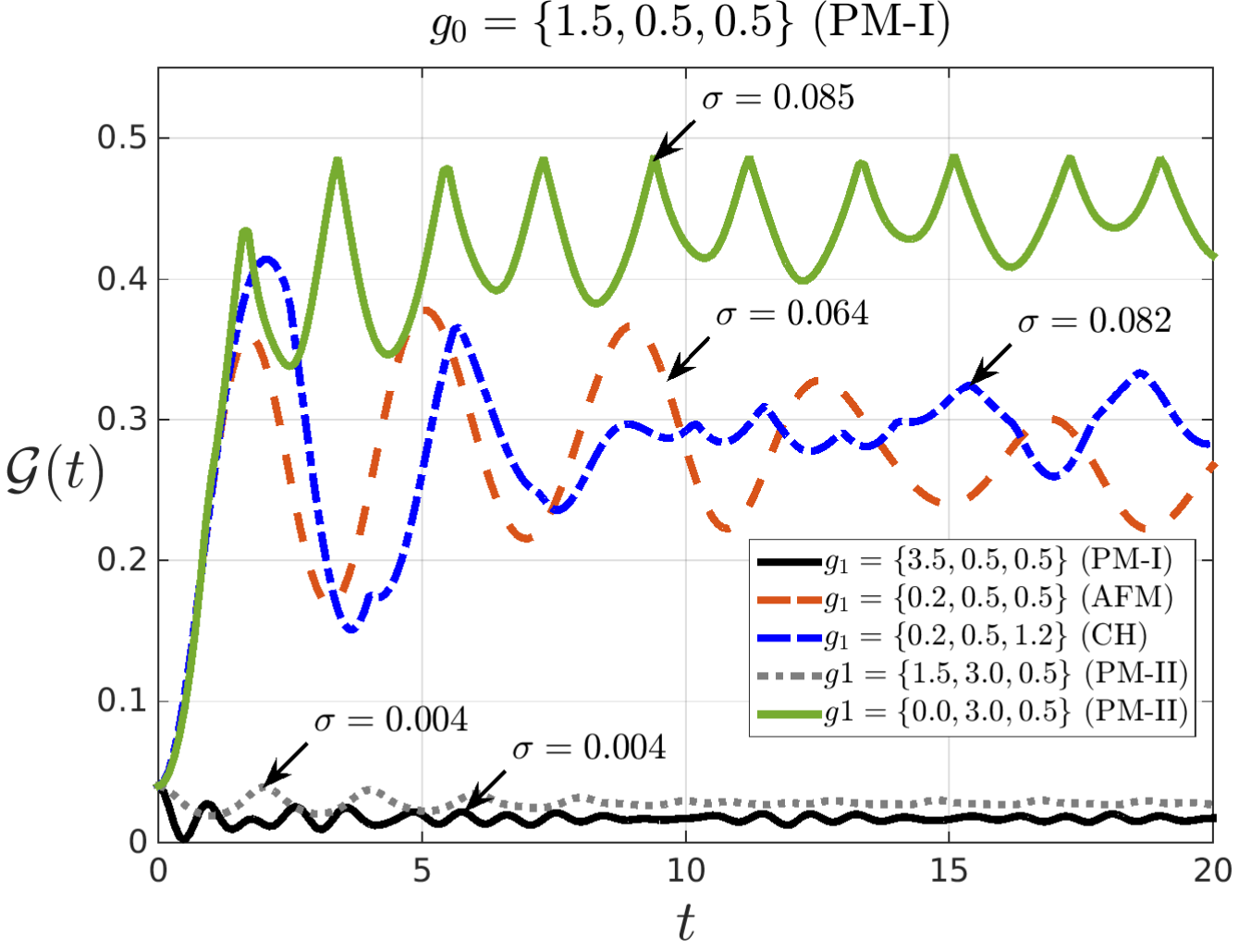}
\caption{(Color online) Time variation of $\mathcal{G}$ after various quenches within and across equilibrium phases. We observe distinctly high amount of fluctuations during the transient period of dynamics for specific regions (corresponding  to DQPT, see Fig. \ref{fig:dpt_ept}) of disorder to order (PM-I to AFM and PM-I to CH), and disorder to disorder (PM-I to PM-II) quenches, irrespective of the size of quench. Apart from these, other quenches show relatively lower fluctuations. Both the axes are dimensionless.}
\label{fig:GGM}
\end{figure}

Although, the evaluation of GGM has a clear prescription, its computation requires finding the maximum eigenvalues of $\binom{N}{1} + \binom{N}{2} + ... + \binom{N}{N/2} \sim 2^N$ - number of matrices which is definitely cumbersome for large $N$. However, from finite size analysis of the DATXY model with $N = 6,8,10$ and $12$, we notice that for almost all times (except the initial response time $\sim \frac{2J}{\hbar}$), the maximal eigenvalue comes either from the single or nearest neighbour two-site reduced density matrices. 
So, we can argue that even for systems with large number of parties, the space consisting of the eigenvalues of single and nearest neighbour two-site reduced density matrices remain the \emph{effective} subspace for computing the GGM. Furthermore, we can exploit the translational invariance of the DATXY model to simplify the scanning space even for the single and two-site reduced density matrices to just $\rho_e,\rho_o,\rho_{eo},\rho_{oe}$.   Here $\rho_{e(o)}$ denotes the single site (reduced) density matrix corresponding to even and odd sites respectively, and $\rho_{eo(oe)}$ is the nearest neighbour two-site density matrices between even-odd (odd-even) sites. Note that $\rho_{eo}$ and $\rho_{oe}$ have the same eigenvalues. Thus, in the thermodynamic limit ($N \rightarrow \infty$), the GGM  can be effectively  computed as
\begin{eqnarray}
\mathcal{G}(|\psi\rangle) \approx  1-\max \{\mu_{\rho_e}^{\max},\mu^{\max}_{\rho_o}, \mu^{\max}_{\rho_{eo}}\},
\end{eqnarray}
Note that even if in some situations the above argument does not remain valid, $\mathcal{G}(\ket{\psi})$ still remains a measure of entanglement for multipartite states, providing an upper bound for GGM. Furthermore, it also remains an LOCC monotone. 
\subsection{Advantages of Multipartite Entanglement as a detector of DQPT}
\label{sec:entmul}

We find that multipartite entanglement, $\mathcal{G}$, can capture DQPT when the quench corresponds to an underlying EQPT involving a disorder to order (PM-I/II $\rightarrow$ AFM/CH) or a disorder to disorder (PM-I(II) $\rightarrow$ PM-II(I)) transition.
In particular, for a quench starting from the disordered phase, the dynamics of $\mathcal{G}$ displays a higher amount of oscillation for a quench that leads to a DQPT compared to those which do not start from such a phase (see Fig. \ref{fig:GGM}).

\begin{figure}
\includegraphics[width=\columnwidth]{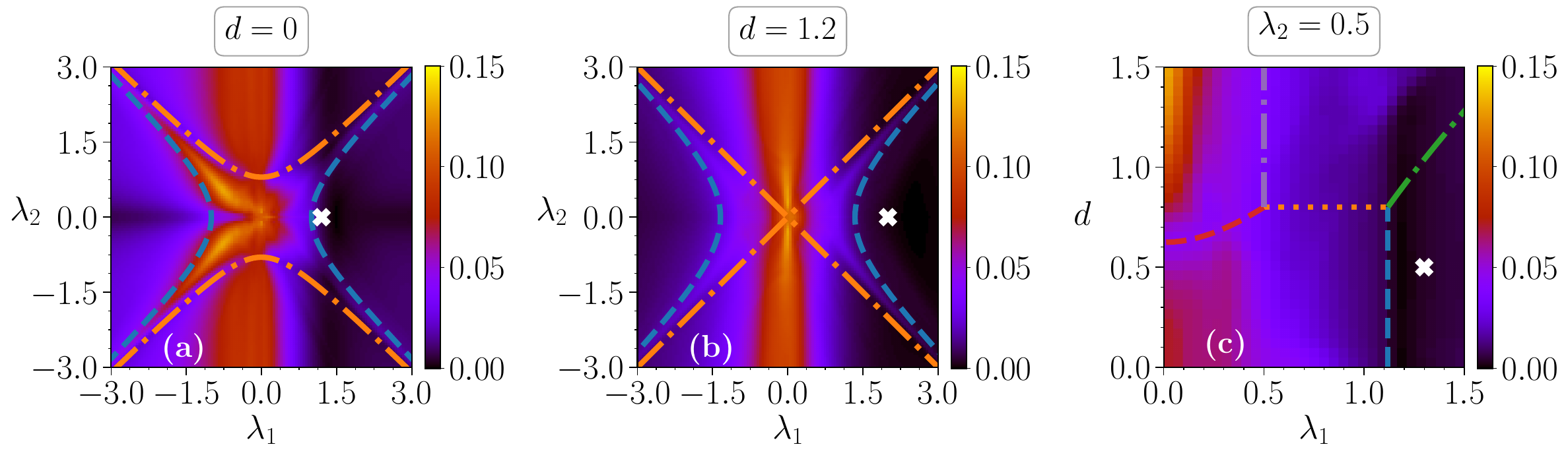}
\caption{ (Color Online) Variation of time averaged standard deviation $\langle\sigma_{\mathcal{G}}(t)\rangle$, for quenches in  ($\lambda_1,\lambda_2,d$)-space with initial choices 
indicated by ``$\times$" as in Fig. \ref{fig:dpt_ept}.  After the quench, we observe the emergence of a specific region in the ($\lambda_1,\lambda_2$)-space in (a), and in ($\lambda_1,d$)-space in (b), which are distinctly  characterized by a high fluctuations, i.e., higher  values of $\langle\sigma_{\mathcal{G}}(t)\rangle$. This region possess a high overlap with that depicted in Fig. \ref{fig:dpt_ept}(a), (b) and (f),  which correspond to DQPTs, obtained by analyzing non-analyticities in the rate function. This substantial overlap, establishes multipartite entanglement as a good detector of DQPT. All the axes are dimensionless.}
\label{fig:dpt_ept_std}
\end{figure}

For a quantitative treatment of the above observation, we estimate the amount of fluctuations in the time dynamics of $\mathcal{G}$ during the transient regime, by computing its time averaged standard deviation, $\langle \sigma_{\mathcal{G}}(t)\rangle$,  defined as
\begin{eqnarray}
\langle \sigma_{\mathcal{G}}(t)\rangle = \frac{\hbar}{J \tau}\int_{0}^{\frac{J\tau}{\hbar}} \sigma_{\mathcal{G}}(t) dt,
\label{eq:fluctuation}
\end{eqnarray}
where $\sigma_{\mathcal{G}}^2(t) =  ({\mathcal{G}^2(t)} - \langle {\mathcal{G}(t)}\rangle)^2$, with $\mathcal{G}(t) = \mathcal{G}(|\psi (t)\rangle)$, and $|\psi(t)\rangle = e^{-i \hat{H}^{(1)} t / \hbar} \ket{\Psi^0}$. Note that for a reasonable average, $\tau$ should be taken to be large, but, on the other hand, it should also be small enough so that the system does not reach a steady state (i.e., $\tau_p < \tau \leq \tau_{st}$ with $\tau_{st}$ being the time at which the system enters a steady state and with $\tau_p$ being the approximate time period, in units of $J/\hbar$, of the oscillations), ensuring that the average is computed in the transient regime even as the effect of small variations in the oscillations are nullified. In our case, we choose $\tau = 20$ (with $\tau_{st} \sim 50$). 
We analyze the variation of $\langle\sigma_{\mathcal{G}}(t)\rangle$ scanning the parameter space ($\lambda_1, \lambda_2, d$) of the DATXY model. It is evident from Figs.  \ref{fig:dpt_ept_std} (a), (b) and (c) (compare with Figs. \ref{fig:dpt_ept} (a), (b) and (f) respectively) that the value of $\langle\sigma_{\mathcal{G}}(t)\rangle$ is substantially larger for quenches that correspond to a DQPT and is
confirmed by a large overlap of such regions with those detected by singularities in the rate function,
 thereby establishing multipartite entanglement, $\mathcal{G}$, as a \emph{good} detector of DQPT.

\begin{figure}
\includegraphics[width=\linewidth]{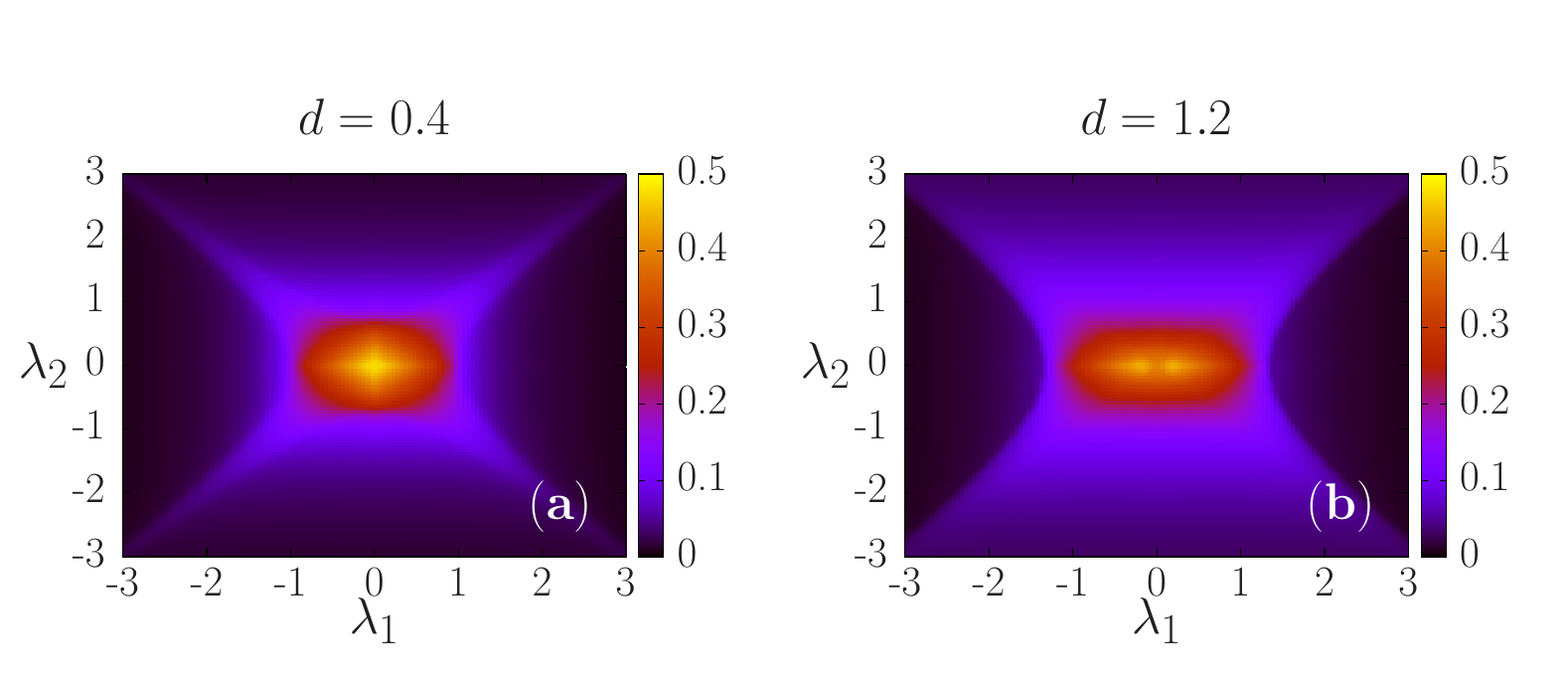}
\caption{The GGM phase portrait of the ground state of the DATXY model for (a) $d=0.5$, and (b) $d=1.2$, with respect to the uniform and alternate fields $\lambda_1$ and $\lambda_2$. The anisotropy parameter $\gamma$ is fixed to $0.8$. All axes are dimensionless.}
\label{fig:ggm}
\end{figure}

%
Despite the advantages offered by multipartite entanglement, its DQPT detection capability is not ubiquitous.  For example, if one starts from an ordered phase, the dynamics of $\mathcal{G}$ cannot detect DQPT. 
Therefore, the time variation of multipartite entanglement considered here, can indicate the presence or absence of DQPTs only when the initial state parameters correspond to a disordered phase. 
A plausible explanation of this asymmetry can be provided by looking at the equilibrium GGM phase portrait which reveals that the ordered phases possess higher values of GGM compared to the disordered phases, see Fig. \ref{fig:ggm}.
We find GGM to be a ``good" detector when the initial state possess low amount of GGM, i.e., when the quench starts from a disordered phase. Note that there is no restriction on the GGM content of the ground state of the driving Hamiltonian. 
We want to stress here that we do not intend to replace the usual markers of many-body phases with GGM, but rather to study the relationship of entanglement with DQPT and highlight that for some quenches, the fluctuations in GGM can actually indicate DQPT. 

\section{Conclusion}
\label{sec:con}
                   
Dynamics of many-body systems reveal qualitative differences depending on the initial and quenched values of system parameters -- known as dynamical quantum phase transition (DQPT). 
In this work, the analysis of DQPT is carried out for the DATXY (alternating field transverse XY model with Dzyaloshinskii-Moriya interaction) model after a sudden quench of system parameters.
We \emph{analytically} found, that in contradistinction to Ising systems, these systems possess non-uniformly spaced critical times (as indicated by zeros of Loschmidt echo) for the quenches which correspond to a DQPT. 
We also proposed a physical quantity based on multipartite entanglement as a good detector of DQPT.


The theory of DQPT, in essence, presents a quantitative formalism to understand the qualitative differences that occur during the dynamics of  many-body systems after quenching of system parameters.
Being intrinsically a feature of the transient regime, the analysis of DQPT is also of practical importance since one does not have to wait until equilibration to observe the relevant physics. Recent experimental realization of DQPT in various physical systems further reinforces the significance of such pragmatic studies.


\acknowledgements
This research was supported in part by the `INFOSYS scholarship for senior students'. 
TC was supported by Quantera QTFLAG project 2017/25/Z/ST2/03029 of National Science Centre (Poland).
Computational work for this study was carried out at the cluster computing facility in the Harish-Chandra Research Institute (http://www.hri.res.in/cluster).

\appendix

\section{Calculation of the rate function}
\label{ap:nonana-ratefn}
The Loschmidt echo, $L(t)$, is then described by the probability associated with this amplitude, i.e., $L(t)= |G(t)|^2$. The rate function associated with $L(t)$, which is analogous to the free energy (per lattice-site) in thermal phase transitions, can be defined as 
\begin{eqnarray}
\mathcal{F}(t) &=& -\lim_{N\rightarrow \infty} \frac{1}{N} \log L(t) \nonumber \\
&=& - \lim_{N \rightarrow \infty} \frac{1}{N} \sum_{p=1}^{N/4} \log |G_p(t)|^2.
\label{eq:rf}
\end{eqnarray} 

To deduce the analytical expressions of the Loschmidt amplitude and the rate function for a general quench from $g_0 \equiv \{\lambda_1(t=0),\lambda_2(t=0),d(t=0)\}$ to  $g_1 \equiv \{\lambda_1(t>0),\lambda_2(t>0),d(t>0)\}$, we introduce the fermionic vector operator $\hat{A_p} = \begin{bmatrix} \hat{a}_p \\ \hat{b}_p  \end{bmatrix}$,  such that we can write $\hat{H_p}$ as $   \hat{H}_p = J \begin{bmatrix}
           \hat{A}^{\dagger T}_{p} & \hat{A}^T_{-p}
             \end{bmatrix} 
           \tilde{H}_p
 \begin{bmatrix}
           \hat{A}_{p}\\ \hat{A}^\dagger_{-p}
            \end{bmatrix},$ with
\small
\begin{eqnarray}
 \tilde{H}_p = 
            \begin{bmatrix}
          (\cos\phi_p + d\sin\phi_p)\hat{\sigma}^x + \Lambda & -i\gamma \sin\phi_p \hat{\sigma}^x \\
          i\gamma \sin \phi_p \hat{\sigma}^x & -(\cos\phi_p-d\sin\phi_p)\hat{\sigma}^x - \Lambda
            \end{bmatrix}, \nonumber \\
\label{eq:hp_tilde}
\end{eqnarray} 
\normalsize  
where $\Lambda = \mbox{Diag}\{\lambda_1-\lambda_2, \lambda_1+\lambda_2\}$. To diagonalize $\tilde{H}_p$ in Eq. \eqref{eq:hp_tilde}, we can perform the Bogoliubov transformation,
\begin{eqnarray}
\begin{bmatrix}
\hat{A}_p \\ \hat{A}^\dagger_{-p}
\end{bmatrix} = 
M_p  \begin{bmatrix} 
    \hat{\Gamma}_p\\
    \hat{\Gamma}^\dagger_{-p}
    \end{bmatrix}
= \begin{bmatrix} 
    U_p & -iV_p\\
    -iV^*_p & U^*_p
    \end{bmatrix}  
    \begin{bmatrix} 
    \hat{\Gamma}_p\\
    \hat{\Gamma}^\dagger_{-p}
    \end{bmatrix},
\end{eqnarray}
with $\hat{\Gamma}_p = \begin{bmatrix} \hat{\eta}^a_p \\ \hat{\eta}^b_p  \end{bmatrix}$, such that $\hat{H}_p$ is diagonal in the Bogoluibov basis,
$\{ {{}\hat{\eta}^a_p } ^ {\dagger}, {{}\hat{\eta}^b_p}^{\dagger}, {\hat{\eta}^a_{-p}}, {\hat{\eta}^b_{-p}} \}$. $U_p$ and $V_p$ are Bogoliubov coefficients (matrices) that diagonalize $\hat{H_p}$ and are functions of $g \equiv \{\lambda_1,\lambda_2, d\} $. The fermionic algebra of $\hat{a}_p$, $\hat{b}_p$, $\hat{\eta}^a_p$, and $\hat{\eta}^b_p$ operators guarantee that the Bogoluibov matrix, $M_p$, is unitary in nature, i.e., $M_p^{-1} = M_p^{\dagger}$.

In order to calculate $\bra{\Psi^0}e^{-iH^{(1)}t/\hbar}\ket{\Psi^0}$, 
 we need to express $\ket{\Psi^0}$ in terms of Bogoluibov operators, $\{\hat{\Gamma}_p(g_1)\}$, that diagonalize $\hat{H}^{(1)}$, where $\hat{H}^{(i)} = \hat{H}(g_i)$, with $g_0$ representing  the parameters of the Hamiltonian whose ground state is the initial state, while $g_1$ defines the parameters of the driving Hamiltonian.   If the operators, $\{\hat{\Gamma}_p(g_0)\}$, diagonalize the initial Hamiltonian, $\hat{H}^{(0)}$, using Eq. \eqref{eq:bogol}, we arrive at the following relation:
 \small
 \begin{eqnarray}
 \begin{bmatrix} 
    \hat{\Gamma}_p(g_0)\\
    \hat{\Gamma}^{\dagger}_{-p}(g_0)
    \end{bmatrix} &=& 
    M^{-1}_p(g_0) M_p(g_1)  
    \begin{bmatrix} 
    \hat{\Gamma}_p(g_1)\\
    \hat{\Gamma}^{\dagger}_{-p}(g_1)
    \end{bmatrix}, \nonumber \\
    &=&\begin{bmatrix} 
    \mathcal{U}_p(g_0, g_1) & -i\mathcal{V}_p(g_0, g_1)\\
    -i\mathcal{V}^*_p(g_0, g_1) & \mathcal{U}^*_p(g_0, g_1)
    \end{bmatrix}  \begin{bmatrix} 
    \hat{\Gamma}_p(g_1)\\
    \hat{\Gamma}^\dagger_{-p}(g_1)
    \end{bmatrix}, \nonumber \\
\end{eqnarray}  
\normalsize
with
 \begin{eqnarray}
    \mathcal{U}_p (g_0, g_1)&=& U^\dagger_p(g_0) U_p(g_1) + V^T_p(g_0)V^*_p(g_1), \nonumber \\
    \mathcal{V}_p (g_0, g_1)&=& U^\dagger_p(g_0)V_p(g_1) - V^T_p(g_0)U^*_p(g_1).
    \end{eqnarray}
    Calculation of $\mathcal{U}_p$ and $\mathcal{V}_p$ matrices entirely depend on the diagonalization of the $4 \times 4$ matrices, $\tilde{H}_p(g_0)$ and $\tilde{H}_p(g_1)$, in Eq. \eqref{eq:hp_tilde}.
Once the matrices, $\mathcal{U}_p$ and $\mathcal{V}_p$, are obtained, we can write the initial state $\ket{\Psi^0}$ as a boundary state composed of zero-momentum modes of $\hat{H}^{(1)}$, which is given by
\begin{eqnarray}
\ket{\Psi^0} = \mathcal{N}^{-1}\exp\Big[i\sum_ {p=1}^{N/4} \hat{\Gamma}^{\dagger T}_p (\mathcal{U}^{-1}_p\mathcal{V}_p) \hat{\Gamma}^\dagger_{-p})\Big]\ket{0},
\label{eq:bs}
\end{eqnarray}
where $\ket{0}$ is the ground state of $\hat{H}^{(1)}$, $\mathcal{N}$ is the normalization constant, and $T$ denotes the transpose of the corresponding operators. If we now assume that  operators $\{\hat{\Gamma}_p(g_1)\}$ can diagonalize the Hamiltonian, $\hat{H}^{(1)}$, in the way, given by
\small
 \begin{eqnarray}
    \hat{H}_p^{(1)} = J 
    \begin{bmatrix}
	\hat{\Gamma}^{\dagger T}_p(g_1) & \hat{\Gamma}^T_{-p}(g_1)
	\end{bmatrix}
	\begin{bmatrix} 
    \hbar \omega^1_p & 0 & 0 & 0 \\
    0 & \hbar \omega^2_p & 0 & 0 \\
    0 & 0 & \hbar\omega^3_p & 0 \\
    0 & 0 & 0 & \hbar \omega^4_p
    \end{bmatrix}  
    \begin{bmatrix} 
    \hat{\Gamma}_p(g_1) \\
    \hat{\Gamma}^\dagger_{-p}(g_1)
    \end{bmatrix}, \nonumber \\
    \end{eqnarray} 
\normalsize
with $\hbar\omega_p^k$, for $k = 1,2,3,4$, being the eigenvalues, then the Loschmidt amplitude, $G(t)=\bra{\Psi^0}e^{-i \hat{H}^{(1)} t / \hbar} \ket{\Psi^0}$, reads  as
\begin{eqnarray}
 G(t) = && \frac{e^{i \frac{J t}{\hbar} \sum_{p=1}^{N/4} (\omega_p^3 + \omega_p^4)}}{\mathcal{N}^{2}}\bra{0}\exp\big[\sum_{p=1}^{N/4} \hat{\Gamma}_{-p}^{T}(g_1)\mathbf{M}_{p} \hat{\Gamma}_p(g_1)\big] \nonumber \\
 &&\times
 \exp\big[\sum_{p=1}^{N/4}  \hat{\Gamma}^{\dagger T}_p(g_1) \mathbf{N}_{p}\hat{\Gamma}_{-p}^{\dagger}(g_1)\big]\ket{0},
 \label{eq:gt_ini}
\end{eqnarray}
where
\begin{eqnarray}
\mathbf{M}_p &=& \begin{bmatrix}
    -i\mathcal{T}^{p*}_{11} & -i\mathcal{T}^{p*}_{21} \\
    -i\mathcal{T}^{p*}_{12} & -i\mathcal{T}^{p*}_{22}
    \end{bmatrix}, \nonumber \\
    \mathbf{N}_p &=& \begin{bmatrix}
    i e^{-i \frac{J t}{\hbar} (\omega^1_p - \omega^3_p)}\mathcal{T}^p_{11} & i e^{-i \frac{J t}{\hbar}(\omega^1_p - \omega^4_p) }\mathcal{T}^p_{12} \\
    i e^{-i \frac{J t}{\hbar}(\omega^2_p - \omega^3_p) }\mathcal{T}^p_{21} & i e^{-i \frac{J t}{\hbar} (\omega^2_p - \omega^4_p}\mathcal{T}^p_{22}
    \end{bmatrix},
    \label{eq:mn}
\end{eqnarray}
with $\mathcal{T}^p_{ij} = (\mathcal{U}^{-1}_p\mathcal{V}_p)_{ij}$. Note that the $\omega_p^k$s are eigenvalues of the matrix $\tilde{H}_p^{(1)}$, and is written as $\hat{H}_p^{(1)} = \hat{A}_p^{\dagger}  \tilde{H}_p^{(1)} \hat{A}_p$, where $\hat{A}_p$ is the column vector,  $(\hat{a}_p, \hat{b}_p, \hat{a}_{-p}^{\dagger}, \hat{b}_{-p}^{\dagger})$, and the $4 \times 4$ matrix, $\tilde{H}_p^{(1)}$, is given as
\begin{widetext}
\small
\begin{equation}
\tilde{H}_p^{(1)} = J
\begin{bmatrix}
(\lambda_1 - \lambda_2) &  (\cos\phi_p + d \sin\phi_p) & 0 & -i \gamma \sin\phi_p \\
(\cos\phi_p + d \sin\phi_p)  & (\lambda_1 + \lambda_2) & -i \gamma \sin\phi_p &0 \\
0 & i\gamma\sin\phi_p & -(\lambda_1 - \lambda_2) &  -(\cos\phi_p - d \sin\phi_p) \\
i \gamma \sin\phi_p & 0 & -(\cos\phi_p - d \sin\phi_p) & -(\lambda_1 + \lambda_2)
\end{bmatrix},
\label{eq:hp_matrix_alt}
\end{equation}
\normalsize
\end{widetext}
with $\phi_p \in [-\pi/2,\pi/2]$.
 Using Eqs. \eqref{eq:gt_ini} and \eqref{eq:mn}, and the prescription developed in \cite{LeClair1995}, we get the Loschmidt amplitude per momentum mode for the DATXY model as

\begin{widetext}
\begin{eqnarray}
G_p(t) &&= e^{i \frac{J t}{\hbar} (\omega_p^3 + \omega_p^4)} \nonumber \\
&&
\resizebox{0.9\hsize}{!}{$\times
\frac{1 
+ e^{-i \frac{J t}{\hbar} (\omega^1_p - \omega^3_p)}|\mathcal{T}^p_{11}|^2 
+ e^{-i \frac{J t}{\hbar} (\omega^2_p - \omega^4_p)} |\mathcal{T}^p_{22}|^2
+ e^{-i \frac{J t}{\hbar}(\omega^1_p - \omega^4_p)}|\mathcal{T}^p_{12}|^2 
+ e^{-i \frac{J t}{\hbar}(\omega^2_p - \omega^3_p) } |\mathcal{T}^p_{21}|^2 
+  e^{-i \frac{J t}{\hbar}(\omega^1_k+\omega^2_k - \omega^3_k-\omega^4_k )} |\mathcal{T}^p_{11}\mathcal{T}^p_{22} - \mathcal{T}^p_{12}\mathcal{T}^p_{21}|^2}
{1 + \vert\mathcal{T}^p_{11} \vert^2 + \vert \mathcal{T}^p_{22} \vert^2 + \vert \mathcal{T}^p_{12} \vert^2 + \vert \mathcal{T}^p_{21} \vert^2 + \vert \mathcal{T}^p_{11}\mathcal{T}^p_{22} - \mathcal{T}^p_{12}\mathcal{T}^p_{21} \vert^2},$} \nonumber \\
\label{eq:LA_p}
\end{eqnarray} 
\normalsize
\end{widetext}
such that $G(t) = \prod_{p=1}^{N/4}G_p(t)$.
Finally, we get the rate function associated with the quench from parameters $g_0$ to $g_1$, in the thermodynamic limit, as
  \begin{eqnarray}
  \mathcal{F}(t) = -\int^{\frac{\pi}{2}}_0 \frac{d\phi_p}{\pi} \log |G_p(t)|.
  \end{eqnarray}
Now, the solutions of $|G_p(t)| = 0$, correspond to the critical times, denoted by $t^*$.  
Clearly, nonanalyticity arises in $\mathcal{F}(t)$, if we can find real solutions $(\phi_p^*, t^*)$ of the transcendental equation
\begin{eqnarray}
 |G_p(t)| = 0,
 \label{eq:eq_t*}
 \end{eqnarray}
 which describes a DQPT with critical time $t^*$. As mentioned before, the matrix, $\mathcal{T}$, and the eigenvalues, $\{\omega_p^k; k=1,2,3,4\}$,
 can be computed easily by diagonalizing $4 \times 4$ matrices, $\tilde{H}_p(g_0)$ and $\tilde{H}_p(g_1)$, which, in turn, allows us to obtain $G_p(t)$, and thus the rate function, $\mathcal{F}(t)$.

\bibliography{bib.bib}

\end{document}